\documentclass[multicol,pra,aps,showpacs,twocolumn,twoside,superscriptaddress]{revtex4}

\usepackage{amsmath,amsfonts,amssymb,latexsym,revsymb,theorem,verbatim}
\usepackage[dvipdfm]{graphicx}
\usepackage{bm}

\newtheorem{definition}{Definition}

\newtheorem{theorem}[definition]{Theorem}

\def\blm{{\boldsymbol \lambda}}

\def\squareforqed{\hbox{\rlap{$\sqcap$}$\sqcup$}}
\def\qed{\ifmmode\squareforqed\else{\unskip\nobreak\hfil
\penalty50\hskip1em\null\nobreak\hfil\squareforqed
\parfillskip=0pt\finalhyphendemerits=0\endgraf}\fi}
\def\endenv{\ifmmode\;\else{\unskip\nobreak\hfil
\penalty50\hskip1em\null\nobreak\hfil\;
\parfillskip=0pt\finalhyphendemerits=0\endgraf}\fi}
\newenvironment{proof}{\noindent \textbf{{Proof.~} }}{\qed}

\def\GU{\mathop{\rm U}\nolimits}

\def\C{{\mathbb{C}}}

\def\E{E_{M,N_1,N_2}}
\def\1R{\rho_1^{\ot N_1}}
\def\2R{\rho_2^{\ot N_2}}
\def\uR{\rho_2^{\ot M + N_2}}
\def\vR{\rho_1^{\ot M + N_1}}
\def\p{p_{ M, N_1, N_2 }}
\def\ap{\overline{p}_{M, N_1, N_2}}
\def\half{\frac{1}{2}}
\def\Tr{\mathop{\rm Tr}\nolimits}
\def\S{ I }
\def\oS{ I_{ N_1 } }
\def\tS{ I_{ N_2 } }
\def\uS{ I_{ M + N_2 } }
\def\vS{ I_{ M + N_1 } }
\def\U{ { \mathcal U } }

\def\kU{ { \mathcal U }_{ [ N - k, k ] } }
\def\oU{ { \mathcal U }_{  N_1 } }
\def\tU{ { \mathcal U }_{ N_2 }}
\def\uU{ { \mathcal U }_{ M + N_2 } }
\def\vU{ { \mathcal U }_{ M + N_1 } }
\def\I{I}

\def\kI{I_{ [ N - k, k ] }}

\def\ktI{I_{ [ N - k, k ] }}
\def\oA{ A_1 }
\def\tA{ A_2 }

\def\tAf{ \frac{ 1 }{ \tA } }
\def\u{u_k}
\def\v{v_k}
\def\w{w_k}
\def\ot{\otimes}
\def\la{\langle}
\def\ra{\rangle}

\def\minn{ \min ( M + N_1, N_2) }
\def\kP{ P_k }
\def\kQ{ Q_k }
\def\ua{ \uparrow }
\def\da{ \downarrow }
\def\al{ \alpha }
\def\be{ \beta }
\def\ga{ \gamma }
\def\hrho{ \hat{ \rho } }

\begin{document}
\title{Changepoint Problem in Quantumn Setting}
\author{Daiki Akimoto}
\address{Graduate School of Information Sciences, Tohoku
University, Aoba-ku, Sendai, 980-8579, Japan}
\author{Masahito Hayashi$^{1,}$}
\email{hayashi@math.is.tohoku.ac.jp}
\address{Centre for Quantum Technologies, National University of
Singapore, 3 Science Drive 2, 117542, Singapore
}

\begin{abstract}
In the changepoint problem, we determine when the distribution observed has changed to another one. We expand this problem to the quantum case where copies of an unknown pure state are being distributed. 
We study the fundamental case, which has only two candidates to choose. 
This problem is equal to identifying a given state with one of the two unknown states when multiple copies of the states are provided. 
In this paper, we assume that two candidate states are distributed independently and uniformly in the space of the whole pure states. The minimum of the averaged error probability is given and the optimal POVM is defined as to obtain it. 
Using this POVM, we also compute the error probability which depends on the inner product. 
These analytical results allow us to calculate the value in the asymptotic case, where this problem approaches to the usual discrimination problem. 
\end{abstract}

\date{\today}



\maketitle

\section{\label{sec:introduction} introduction}

The changepoint problem, which is studied in many fields, originally arose out of considerations of quality control. When a process is ``in control," products are produced according to some rule. At the unknown point the process jumps ``out of control" and ensuing products are produced according to another rule. It is necessary to determine the changepoint.

In the classical case, we observe sequentially a discrete series of independent observations $X_0, X_1, \dots$ 
whose distribution possibly changes at an unknown point in time. 
It is assumed that
independent random variables
$X_0, X_1,\dots, X_{\nu - 1}$ are each distributed according to a distribution and 
the remaining independent random variables
$X_{\nu}, X_{\nu + 1}, \dots$ are each distributed according to another distribution. Our purpose is to detect the changepoint ``$\nu$" \cite{changepoint}. 

We extend this problem to the quantum setting where copies of an unknown pure state are being distributed in discrete time. 
Consider now the device distributing copies of an unknown pure state. 
This device has the unknown changepoint and distributes copies of another unknown pure state after the changepoint.
Our goal is estimating the changepoint by observing all copies the device distributed.

In this paper, we deal with the fundamental case, that is, we have only two candidates for the changepoint.
We assume that the device distributes 
unknown pure states $\rho_t$ on the $d$-dimensional space
for the discrete time $t=0,1,2,\ldots, t_3$.
The state $\rho_t$ is changed at the changepont $t_c$.
That is,
the states $\rho_0, \ldots, \rho_{t_c-1}$ are identical,
and 
the other states $\rho_{t_c}, \ldots, \rho_{t_3}$ 
are also identical.
Further, we assume that there are two candidates of the changepoint $t_1$ and $t_2$.

\begin{figure}[t]
\begin{center}
\scalebox{1.0}{\includegraphics[scale=0.28]{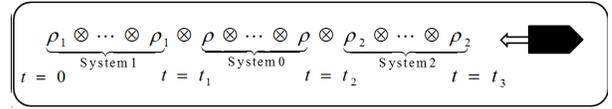}}
\end{center}
\caption{Systems $1$, $0$, and $2$}
\label{fh1}
\end{figure}

In order to analyze this problem,
we introduce System $1$, System $0$ and System $2$ to denote the systems 
that are the composite systems corresponding to 
the time period $0 \le t < t_1, t_1 \le t < t_2$ and $t_2 \le t \le t_3$, respectively, as is explained in Fig. \ref{fh1}. 
Our task is then choosing the correct changepoint $t_c$
by using all three systems. 
This problem is equal to 
decide whether the state in System $0$ coincides with the state in System $1$ or the state in System $2$. 

We derive the optimal POVM and the minimum of the averaged error probability. 
This minimum probability is already obtained 
when the numbers of copies of states in System $1$ and System $2$ are the same. 
Sentis et al. studied this problem in the qubits case, i.e., in the case of $d=2$ \cite{Bagan} and A. Hayashi et al. discussed it when System 0 only has one copy \cite{Hayashi}. 
Our result concerns the general case that has no restriction for the numbers of copies and the dimension in systems.

We also compute the error probability under the application of our optimal POVM, 
not averaged, 
which depends on the inner product of two states in System $1$ and System $2$.
As the numbers of copies in three systems approach infinity, 
this error probability clearly approaches $0$ exponentially unless the inner product is $1$. 
Hence, the convergence speed can be measured by the exponential decreasing rate. 
The exponential decreasing rate seems related to the quantum Chernoff bound, 
which gives the optimal exponential decreasing rate 
of two state discrimination when a large number of copies of the unknown state are available \cite{Chernoff},
while this relation has not been discussed.
In this paper, using the above analytical result, 
we calculate the exponential decreasing rate
and clarify the relationship with the quantum Chernoff bound.

The paper is organized as follows. 
In the next Section we give the optimal strategy for the minimum averaged error probability. 
The optimal POVM is described by using the representation theory for easy calculation in the following sections. 
In Section III we obtain the minimum error probability represented as a function of the number of copies in each three systems and the dimension of the state space. 
Using the optimal POVM, in Section IV we compute the error probability which depends on the inner product of two states in System $1$ and System $2$.
In Section V we consider the asymptotic behaviors of the minimum averaged error probability in several scenarios.
In Section VI,
we finally compute the convergence speed of the error probability. Some brief conclusions follow and we end up with a technical appendix.
\section{The Optimal POVM}

In order to treat the decision problem of the changepoint,
we denote the numbers $t_1$, $t_2-t_1$, $t_3-t_2+1$ by
$N_1$, $M$, $N_2$, respectively.
Then, we have $N_1$ ($N_2$) copies of unknown pure states in System $1$ ($2$). 
In the following, we denote 
the unknown state on System $1$ ($2$)
by $\rho_1$ ($\rho_2$), which is a pure state on 
the $d$ dimensional vector space $\C^d$.
System 0 has $M$ copies of the unknown state $\rho$ that is guaranteed to be either one of $\rho_1$ and $\rho_2$. 
Note that we assume that $N_1 \le N_2$ which loses no generality of this problem. Our purpose is to identify the state $\rho$ with one of the two states by using all systems. 
This is equal to distinguishing two states, $\1R \ot \rho_1^{\ot M} \ot \2R$ and $\1R \ot \rho_2^{\ot M} \ot \2R$, which are assumed to occur with equal probability.

In this decision problem,
we apply two-valued POVM $\{E_1,E_2\}$, in which
$E_1$ ($E_2$) corresponds to the decision $t_c=t_1$ ($t_c=t_2$).
Since $E_1=I-E_2$, 
our POVM can be described by a Hermitian matrix $0 \le E_2 \le I$,
where $I$ is the unit matrix. 
Then, the error probability is defined as
\begin{align}
&\p ( \rho_1, \rho_2, E_2 ) \nonumber \\
\equiv & \half  \Tr [ ( \1R \ot \uR ) E_2  ]
+ \half \Tr [ ( \vR \ot \2R ) E_1] \nonumber \\
= & \half  \Tr [ ( \1R \ot \uR ) E_2  ]\nonumber \\
&\hspace{12ex}+ \half \Tr [ ( \vR \ot \2R ) (I-E_2)] 
\label{eq:p}.
\end{align}

Now, we assume that $\rho_1$ and $\rho_2$ are independently distributed according to the unitary invariant distribution $\mu_{\Theta_d}$
on the set $\Theta_d$ of pure states on 
the $d$ dimensional vector space $\C^d$.
By using the POVM $\{ I-E_2, E_2 \}$, we can define the averaged error probability as
\begin{equation}
\begin{split}
&\ap ( E_2 ) \\
&~\equiv \int_{ \Theta_d } \int_{ \Theta_d } \p ( \rho_1, \rho_2, E_2 ) \mu_{ \Theta_d } ( d \rho_1 ) \mu_{ \Theta_d } ( d \rho_2 ) \label{eq:ap}.
\end{split}
\end{equation}
Here it is very helpful to use the following formula for the integral of the tensor product of $L$ identically prepared pure states \cite{Hayashi}:
\begin{equation}
\int_{ \Theta_d } \sigma^{ \ot L } \mu_{ \Theta_d } ( d \sigma ) = \frac{\S_L}{\Tr[ \S_L ]} \label{eq:Schur},
\end{equation}
where $\sigma$ is a pure state and $\S_L$ is the projector onto the totally symmetric subspace of $( \C^d )^{ \ot L }$.

By using this formula, the averaged error probability reads
\begin{equation}
\begin{split}
&\ap ( E_2 )\\
&~= \half \left[ 1 + \Tr[ ( \frac{\oS \ot \uS}{\oA} - \frac{\vS \ot \tS}{\tA} ) E_2 ] \right] \label{eq:ap1},
\end{split}
\end{equation}
where $A_1$ and $A_2$ are defined as follows:
\begin{equation}
\begin{split}
\oA &\equiv \Tr[ \oS ] \Tr[ \uS ] \\
&= \binom{N_1 + d - 1}{d - 1} \binom{M + N_2 + d - 1}{d - 1} \label{eq:oA}.
\end{split}
\end{equation}
\begin{equation}
\begin{split}
\tA &\equiv \Tr[ \vS ] \Tr[ \tS ] \\
&= \binom{M + N_1 + d - 1}{d - 1} \binom{N_2 + d - 1}{d - 1} \label{eq:tA}.
\end{split}
\end{equation}
Note that $A_1 \le A_2$, and the equation holds if and only if $N_1 = N_2$.  

Eq. (\ref{eq:ap1}) guarantees that
the optimal strategy to minimize the averaged error probability
is given by the Hermitian matrix
\begin{equation}
\E \equiv \{ \frac{\oS \ot \uS}{\oA} - \frac{\vS \ot \tS}{\tA} < 0 \} \label{eq:E},
\end{equation}
where $\{ A < 0 \}$ represents a projector onto the eigenspaces with negative eigenvalues of $A$. 
That is, plugging Eq. (\ref{eq:E}) into Eq. (\ref{eq:ap1}), one obtains the minimum averaged error probability. 

In order to compute the minimum averaged error probability, 
we deform the expression of the optimal POVM 
by using the tensor product representation of the unitary group $\GU ( d )$.

Any irreducible representation of the unitary group $\GU ( d )$
is characterized by a Young diagram $\displaystyle \blm = [ \lambda_1, \lambda_2, \dots, \lambda_d ]$
and is denoted by $\U_{ \blm }$.
We use the shorthand notations $\lambda_1$ and $[ \lambda_1, \lambda_2 ]$ to denote 
$[ \lambda_1, 0, 0, \dots, 0 ]$ and $[ \lambda_1, \lambda_2, 0, 0, \dots, 0 ]$. 
Note that $\U_{  L }$ means the totally symmetric subspace of $( \C^d )^{ \ot L }$. The dimension of $\U_{ [ \lambda_1, \lambda_2 ] }$ is given as:
\begin{equation}
\dim \U_{ [ \lambda_1, \lambda_2 ] }= \frac{ ( \lambda_1 + d - 1 ) ! ( \lambda_2 + d - 2 ) ! ( \lambda_1 - \lambda_2 + 1) }{ ( d - 1 ) ! ( d - 2 ) ! ( \lambda_1 + 1 ) ! \lambda_2 ! } \label{eq:dimU}.
\end{equation}

In our problem, the total system size is $N \equiv M + N_1 + N_2$,
and the total tensor product space $( \C^d )^{ \ot N }$
can be decomposed to 
\begin{equation}
( \C^d )^{ \ot N } = \oplus_{ \blm } {\mathcal U}_{ \blm } \ot {\mathcal V}_{ \blm } \label{eq:decomposition},
\end{equation}
where ${\mathcal V}_{ \blm }$ corresponds to the multiplicity of the irreducible space ${\mathcal U}_{ \blm }$.

Since the tensor product space $( \C^d )^{ \ot N }$ contains
the two subspaces $\oU \ot \uU$ and $\vU \ot \tU$ without multiplicity,
these two subspaces have the form:
\begin{align}
&\oU \ot \uU = \oplus_{ k = 0 }^{ N_1 } {\mathcal U}_{ [ N - k, k ] } \ot \C | \u \ra \label{eq:uU}, \\ 
&\vU \ot \tU = \oplus_{ k = 0 }^{ \minn } {\mathcal U}_{ [ N - k, k ] } \ot \C | \v \ra \label{eq:vU}
\end{align}
by using two normalized vectors:
\begin{align}
&| \u \ra \in {\mathcal V}_{ [ N - k, k ] } ~~( 0 \le k \le N_1 ) \label{eq:u}, \\
&| \v \ra \in {\mathcal V}_{ [ N - k, k ] } ~~( 0 \le k \le \minn ) \label{eq:v},
\end{align}
satisfying $\la \u | \v \ra \ge 0$.
Since the dimension of ${\mathcal V}_{  N }$ is one,
the relation $| u_0 \ra = | v_0 \ra$ holds.

Letting $\I_{ \blm }$ be the projector onto the space ${\mathcal U}_{ \blm }$,
one obtains from the equations (\ref{eq:uU}) and (\ref{eq:vU})
\begin{align}
& \oS \ot \uS = \sum_{ k = 0 }^{ N_1 } \kI \ot | \u \ra \la \u | \label{eq:uS}, \\
& \vS \ot \tS = \sum_{ k = 0 }^{ \minn } \kI \ot | \v \ra \la \v | \label{eq:vS}.
\end{align}
Here, we note that the ranges of summations are different from each other. Using these equations, one has
\begin{equation}
\begin{split}
& \frac{\oS \ot \uS}{\oA} - \frac{\vS \ot \tS}{\tA} \\
&~= \sum_{ k = 0 }^{ N_1 } \kI \ot \left( \frac{| \u \ra \la \u |}{\oA} - \frac{| \v \ra \la \v |}{\tA} \right) \\
&~~~~~- \sum_{ k = N_1 + 1 }^{ \minn } \kI \ot \frac{| \v \ra \la \v |}{\tA} \label{eq:SS}.
\end{split}
\end{equation}
Since $| \u \ra$ and $| \v \ra$ are linearly independent in the range $1 \le k \le N_1$ (we show in Eq. (\ref{eq:ip1})), 
there exists only one negative eigenvalue of $\frac{| \u \ra \la \u |}{\oA} - \frac{| \v \ra \la \v |}{\tA}$.
Using the normalized eigenvector $| \w \ra \in {\mathcal V}_{ [ N - k, k ] }$ 
with this eigenvalue,
we therefore can write the optimal POVM as
\begin{equation}
\begin{split}
\E = \sum_{ k = 1 }^{ N_1 } &\kI \ot | \w \ra \la \w | \\
&+ \sum_{ k = N_1 + 1 }^{ \minn } \kI \ot | \v \ra \la \v | \label{eq:E1}.
\end{split}
\end{equation}

\section{The Minimum Averaged Error Probability}

In this section, we will compute the minimum averaged error probability. Plugging equations (\ref{eq:SS}) and (\ref{eq:E1}) into Eq. (\ref{eq:ap1}) as $E_2 = \E$, one obtains
\begin{equation}
\begin{split}
&\ap ( \E )  \\
&= \half \Bigg[ \sum_{ k = 1 }^{ N_1 } \Tr [ \kI ] \la \w | \Bigg( \frac{| \u \ra \la \u |}{\oA} - \frac{| \v \ra \la \v |}{\tA} \Bigg) | \w \ra \\
&~~~~- \tAf \sum_{ k = N_1 + 1 }^ { \minn } \Tr [ \kI ] + 1 \Bigg] \label{eq:ae2}.
\end{split}
\end{equation}

When arbitrary two real non-zero constants $C_1$ and $C_2$ and 
two linearly independent normalized vectors $| a \ra$ and $|b \ra$ are given, 
the unique negative eigenvalue of $\frac{ | a \ra \la a | }{ C_1 } - \frac{| b \ra \la b |}{ C_2 }$ is given by
\begin{equation}
\frac{ C_2 - C_1 - \sqrt{ ( C_2 - C_1 )^2 + 4 C_1 C_2 ( 1 - | \la a | b  \ra |^2 ) } }{ 2 C_1 C_2 } \label{eq:general}.
\end{equation}
Therefore, 
the eigenvector $| \w \ra$ associated with the negative eigenvalue satisfies 
\begin{equation}
\begin{split}
& \la \w | \left( \frac{| \u \ra \la \u |}{\oA} - \frac{| \v \ra \la \v |}{\tA} \right) | \w \ra \\
&~= \frac{ \tA - \oA - \sqrt{ ( \tA - \oA )^2 + 4 \oA \tA ( 1 - | \la \u | \v \ra |^2 ) } }{ 2 \oA \tA } \label{eq:AA}.
\end{split}
\end{equation}
We also obtain the following equations:
\begin{align}
\Tr [ \kI ] & = \dim \kU \label{eq:dimU1},\\ 
\sum_{ k = 0 }^{ N_1 } \Tr [ \kI ] & = \oA \label{eq:Tsum1}, \\
\sum_{ k = 0 }^{ \minn } \Tr [ \kI ] & = \tA \label{eq:Tsum2}.
\end{align}
Using these equations, we can write the minimum averaged error probability as
\begin{equation}
\begin{split}
&\ap ( \E ) = \\
&~\frac{ 1 }{ 4 } \Bigg[ \frac{ \oA + \tA }{ \tA } \\
&- \sum_{ k = 0 }^{ N_1 } \frac{ \dim \kU }{ \oA \tA } \sqrt{ ( \tA - \oA )^2 + 4 \oA \tA ( 1 - | \la \u | \v \ra |^2 ) } \Bigg] \label{eq:ap3}.
\end{split}
\end{equation}

Our remained task is to calculate the inner product $\la \u | \v \ra$. 
When we denote 
the highest weight vector of the space ${\mathcal U}_{ [ N - k, k ] }$
by $| [ N - k, k ]^d \ra$,
$\la \u | \v \ra$ is equal to the inner product of $| [ N - k, k ]^d \ra | \u \ra$ and $| [ N - k, k ]^d \ra | \u \ra$. 
We can assume $d = 2$ without loss of generality
since the inner product does not depend on the dimension. 
Let us fix some notations as follows:
\begin{equation}
\begin{split}
&\mu_0 \equiv \frac{ M }{ 2 }, \mu_1 \equiv \frac{ N_1 }{ 2 }, \mu_2 \equiv \frac{ N_2 }{ 2 }, \\
&\mu_{ 01 } \equiv \frac{ M + N_1 }{ 2 }, \mu_{ 02 } \equiv \frac{ M + N_2 }{ 2 }, \mu \equiv \frac{ N }{ 2 } - k \label{eq:mu}.
\end{split}
\end{equation}
Using Wigner's 6j-function \cite{Messiah}, we then can write
\begin{equation}
\begin{split}
&\la \u | \v \ra \\
=& ( -1 )^{ \mu_0 + \mu_1 + \mu_2 + \mu } \sqrt{ ( 2 \mu_{ 01 } + 1 ) ( 2 \mu_{ 02 } + 1 ) } 
\begin{Bmatrix} \mu_1 & \mu_0 & \mu_{ 01 } \\ \mu_2 & \mu  & \mu_{ 02 } \end{Bmatrix} \label{eq:ip}.
\end{split}
\end{equation}
Moreover, Wigner's 6j-function can be computed as
\begin{equation}
\begin{split}
&\begin{Bmatrix} \mu_1 & \mu_0 & \mu_{ 01 } \\ \mu_2 & \mu  & \mu_{ 02 } \end{Bmatrix} \\
&= \frac{  ( -1 )^{ \mu_0 + \mu_1 + \mu_2 + \mu } }{ \sqrt{ ( 2 \mu_{ 01 } + 1 ) ( 2 \mu_{ 02 } + 1 ) } } 
\sqrt{ \frac{ \binom{ N_1 }{ k } \binom{ N_2 }{ k } }{ \binom{ M + N_1 }{ k } \binom{ M + N_2 }{ k } } } \label{eq:6j}
\end{split}
\end{equation}
(Appendix 1). Thus, one obtains
\begin{equation}
\la \u | \v \ra = \sqrt{ \frac{ \binom{ N_1 }{ k } \binom{ N_2 }{ k } }{ \binom{ M + N_1 }{ k } \binom{ M + N_2 }{ k } } } \label{eq:ip1},
\end{equation}
and in order to denote this value we use the notation $\phi_k$ satisfying
\begin{equation}
\cos \phi_k = \sqrt{ \frac{ \binom{ N_1 }{ k } \binom{ N_2 }{ k } }{ \binom{ M + N_1 }{ k } \binom{ M + N_2 }{ k } } } = \la \u | \v \ra \label{eq:phi}.
\end{equation}

Therefore, the minimum averaged error probability can be written as
\begin{align}
\begin{split}
&\ap ( \E ) \\ 
&~= \frac{ 1 }{ 4 } \Bigg[ \frac{ \oA + \tA }{ \tA } \\
&~~- \sum_{ k = 0 }^{ N_1 } \frac{ \dim \kU }{ \oA \tA }  \sqrt{ ( \tA - \oA )^2 + 4 \oA \tA \sin^2 \phi_k } \Bigg] \label{eq:ap4}.
\end{split}
\end{align}
In the case of $d = 2$, 
since
\begin{align*}
&A_1 = (N_1 + 1) (M + N_2 + 1) \\
&A_2 = (M + N_1 + 1) (N_2 + 1) \\
&A_2 - A_1 = M (N_2 - N_1) \\
&\dim \U_{[N - k, k]} = M + N_1 + N_2 - 2 k + 1,
\end{align*}
(\ref{eq:ap4}) is calculated to the following way.
\begin{widetext}
\begin{equation}
 \begin{split}
 & \ap ( \E ) = \frac{1}{4} \Bigg[ 1 + \frac{(N_1 + 1) (M + N_2 + 1)}{(M + N_1 + 1) (N_2 + 1)} 
 - \sum_{k = 0}^{N_1} \frac{M + N_1 + N_2 - 2 k + 1}{(N_1 + 1) (M + N_2 + 1) (M + N_1 + 1) (N_2 + 1)} \times \\
 &\sqrt{M^2 (N_2 - N_1)^2 + 4 (N_1 + 1) (M + N_2 + 1) (M + N_1 + 1) (N_2 + 1) 
 \left[ 1 - \left( \frac{N_1 ! (M + N_1 - k)!}{(M + N_1)! (N_1 - k)!} \right) \left( \frac{N_2 ! (M + N_2 - k)!}{(M + N_2)! (N_2 - k)!} \right) \right]} \Bigg]
 \end{split}
\label{our-eq}
\end{equation}
\end{widetext}

When $N_1 = N_2$,  the equation $\oA = \tA$ holds and this probability is concretely computed as
\begin{equation}
\begin{split}
&\overline{ p }_{ M, N_1, N_1 } ( E_{ M, N_1, N_1 } ) \\
&~= \half \Biggl[ 1 - \frac{ ( d - 1 )  N_1 ! ( M + N_1 ) ! }{ ( N_1 + d - 1 ) ! ( M + N_1 + d - 1 ) ! } \\
&~~\times \sum_{ k = 0 }^{ N_1 } ( M + 2 N_1 - 2 k + 1 ) \\
&~~~\times \frac{  ( M + 2 N_1 - k + d - 1 ) ! ( k + d - 2 ) ! }{ ( M + 2 N_1 - k + 1 ) ! k ! } \\
&~~~\times \sqrt{ 1 - \left( \frac{ N_1 ! ( M + N_1 - k ) ! }{ ( M + N_1 ) ! ( N_1 -k ) ! } \right) ^2 } \Biggr] \label{eq:apeq}.
\end{split}
\end{equation}

Moreover, plugging $d = 2$ into this equation, we have
\begin{equation}
\begin{split}
&\overline{ p }_{ M, N_1, N_1 } ( E_{ M, N_1, N_1 } ) \\
&~= \half \Bigg[1 - \sum_{k = 0}^{N_1} \frac{M + 2 N_1 - 2 k + 1}{(N_1 + 1)(M + N_1 + 1)} \\
&~~\times \sqrt{1 - (\frac{N_1 ! (M + N_1 - k)!}{(M + N_1)! (N_1 - k)!})^2} \Bigg] \label{eq:apeq2}.
\end{split}
\end{equation}

This result coincides with the result by Sentis et al.\cite{Bagan}\footnote{Sentis et al.\cite{Bagan} also treated
the case of $N_1\neq N_2$.
In their derivation, 
they simply apply (10) of \cite{Bagan} with the inner product (\ref{eq:ip1}) 
to each irreducible component.
It can be applied only when $A_1=A_2$.
However, in the case of $N_1\neq N_2$,
the equation $A_1=A_2$ does not hold.
Hence, their result (A2) is valid only when $N_1=N_2$.
In our derivation, we apply the formula (\ref{eq:AA}) instead of their (10).
In fact, their result (A2) is different from our result (\ref{our-eq}).}.

\section{The Error Probability with the Optimal POVM}

In the previous section, we obtained the averaged error probability when the two candidate states are distributed independently. 
However, the unknown states $\rho_1$ and $\rho_2$ do not necessarily obey the uniform distribution.
In order to treat the performance of our Optimal POVM in a more general setting,
we consider the error probability for the given two pure states $\rho_1$ and $\rho_2$ 
when the optimal POVM is applied.
This error probability depends on 
the inner product of two candidates, i.e., $q \equiv \Tr [ \rho_1 \rho_2 ]$. 
In the following, we calculate the error probability given by Eq. (\ref{eq:p}) in the case of the optimal POVM $\{  I - \E,\E \}$. 

\begin{theorem}
The error probability with the optimal POVM $\{  I - \E,\E \}$ can be written as
\begin{equation}
\begin{split}
&\p ( \rho_1, \rho_2, \E ) \\ 
&~= \frac{ 1 }{ 4 } \sum_{ k = 0 }^{ N_1 } \Bigg[ \kP + \kQ  \\
&~- \frac{ ( \tA - \oA ) ( \kP - \kQ ) + 2 \sin^2 \phi_k ( \oA \kP + \tA \kQ ) }{ \sqrt{ ( \tA - \oA )^2 + 4 \oA \tA \sin^2 \phi_k } } \Bigg] \label{eq:p1},
\end{split}
\end{equation}
where $P_k$ and $Q_k$ is given as follows:
\begin{equation}
\begin{split}
&\kP \equiv \frac{ ( N - 2 k + 1 ) N_1 ! ( M + N_2 ) ! }{ ( N - k + 1 ) ! k ! } \\
&~\times\sum_{ l = 0 }^{ N_1 - k } \binom{ N_1 - k }{ l } \binom{ M + N_2 - k + l }{ l } q^l ( 1 - q )^{ N_1 - l } \label{eq:P},
\end{split}
\end{equation}
\begin{equation}
\begin{split}
&\kQ \equiv \frac{ ( N - 2 k + 1 ) ( M + N_1 ) ! N_2 ! }{ ( N - k + 1 ) ! k ! } \\
&~\times \sum_{ l = 0 }^{ N_2 - k } \binom{ M + N_1 - k + l }{ l } \binom{ N_2 - k }{ l }  q^l ( 1 - q )^{ N_2 - l } \label{eq:Q}.
\end{split}
\end{equation}
Here we have defined $0^0 \equiv 1$.

When $N_1 = N_2$, we can write
\begin{equation}
\begin{split}
&p_{ M, N_1, N_1 } ( \rho_1, \rho_2, E_{ M, N_1, N_1 } ) \\
&~= \half \Bigg[ 1 - \sum_{ k = 0 }^{ N_1 } \frac{ ( M + 2 N_1 - 2 k + 1 ) N_1 ! ( M + N_1 ) ! }{ ( M + 2 N_1 - k + 1 ) ! k ! } \\
&~~\times \sqrt{ 1 - \left( \frac{ N_1 ! ( M + N_1 - k ) ! }{ ( M + N_1 ) ! ( N_1 -k ) ! } \right)^2 } \\
&~~\times \sum_{ l = 0 }^{ N_1 - k } \binom{ N_1 - k }{ l } \binom{ M + N_1 - k + l }{ l } q^l ( 1 - q )^{ N_1 - l } \Bigg] \label{eq:peq}.
\end{split}
\end{equation}
\end{theorem}

\begin{proof}
Let us start by defining the notation $P_k$ and $Q_k$ as
\begin{equation}
\begin{split}
&\kP \equiv \Tr [ \kI \ot | \u \ra \la \u | ( \1R \ot \uR ) ] \\
&~~~~~~~~~~~~~~( 0 \le k \le N_1 ) \label{eq:P0},
\end{split}
\end{equation}
\begin{equation}
\begin{split}
&\kQ \equiv \Tr [ \kI \ot | \v \ra \la \v | ( \vR \ot \2R ) ] \\
&~~~~~~~~~~~~~~( 0 \le k \le \minn ) \label{eq:Q0}.
\end{split}
\end{equation}
We note that the followin equation holds:
\begin{equation}
\sum_{ k = 0 }^{ N_1 } \kP = \sum_{ k = 0 }^{ \minn } \kQ = 1 \label{eq:sumPQ}.
\end{equation}

Since an arbitrary pure state $\sigma$ satisfies
\begin{equation}
\S_L^d \sigma^{ \ot L } \S_L^d = \sigma^{ \ot L } \label{eq:general1},
\end{equation}
one obtains
\begin{equation}
\begin{split}
&\Tr[ ( \1R \ot \uR ) \E ] \\
&~~= \Tr [ ( \oS \ot \uS ) ( \1R \ot \uR ) ( \oS \ot \uS ) \\
&~~~~~~~~~~~~~\times \E ] \label{eq:TrE}.
\end{split}
\end{equation}
Plugging equations (\ref{eq:uS}) and (\ref{eq:E1}) into Eq. (\ref{eq:TrE}), one has
\begin{align}
& \Tr[ ( \1R \ot \uR ) \E ] = \sum_{ k = 1 }^{ N_1 } | \la \v | \w \ra |^2 \kP \label{eq:P1}. \\
\intertext{In the same way, using equations (\ref{eq:vS}) and (\ref{eq:TrE}), we obtain}
\begin{split}
&\Tr[ ( \vR \ot \2R ) \E ] \\
&~~~~~~~~= \sum_{ k = 1 }^{ N_1 } | \la \v | \w \ra |^2 \kQ + \sum_{ k = N_1 + 1 }^{ \minn } \kQ \label{eq:Q1}.
\end{split}
\end{align}

When two arbitrary normalized and linearly independent vectors$| a \ra, | b \ra$ 
and two positive real numbers $C_1, C_2 >0$ are given,
the normalized eigenvector $| - \ra$ with the unique negative eigenvalue of $\frac{ | a \ra \la a | 
}{ C_1 } - \frac{| b \ra \la b |}{ C_2 }$ ~
satisfies the following equations:
\begin{align}
| \la a | - \ra |^2 &= \half \left[ 1 - \frac{ C_2 - C_1 + 2 C_1 ( 1 - | \la a | b \ra |^2 ) }{ \sqrt{ ( C_2 - C_1 )^2 + 4 C_1 C_2 ( 1 - | \la a | b \ra |^2 ) } } \right] \label{eq:generala},\\
| \la b | - \ra |^2 &= \half \left[ 1 - \frac{ C_2 - C_1 - 2 C_2 ( 1 - | \la a | b \ra |^2 ) }{ \sqrt{ ( C_2 - C_1 )^2 + 4 C_1 C_2 ( 1 - | \la a | b \ra |^2 ) } } \right] \label{eq:generalb}.
\end{align}
Applying equations (\ref{eq:generala}) and (\ref{eq:generalb}) to
the case of $| a \ra= | \u \ra, | b \ra=| \v \ra$,
we have
\begin{align}
| \la \u | \w \ra |^2 &= \half ( 1 - \frac{ \tA - \oA + 2 \oA \sin^2 \phi_k }{ \sqrt{ ( \tA - \oA )^2 + 4 \oA \tA \sin^2 \phi_k } } ) \label{eq:uw},\\
| \la \v | \w \ra |^2 &= \half ( 1 - \frac{ \tA - \oA - 2 \tA \sin^2 \phi_k }{ \sqrt{ ( \tA - \oA )^2 + 4 \oA \tA \sin^2 \phi_k } } ) \label{eq:vw}.
\end{align}
for $1 \le k \le N_1$,

Using these equations, we can write the error probability as
\begin{equation}
\begin{split}
&\p ( \rho_1, \rho_2, \E ) \\
&~= \half \Bigg[ 1 + \Tr[ ( \1R \ot \uR ) \E ] \\
&~~~~~~~~~~~~~~~- \Tr[ ( \vR \ot \2R ) \E ] \Bigg] \\
&~= \half \Bigg[ 1 + \sum_{ k = 1 }^{ N_1 } | \la \u | \w \ra |^2 \kP - \sum_{ k = 1 }^{ N_1 } | \la \v | \w \ra |^2 \kQ \\
&~~~~~~~~~~~~~~~- \sum_{ k = N_1 + 1 }^{ \minn } \kQ \Bigg] \\
&~= \frac{ 1 }{ 4 } \sum_{ k = 0 }^{ N_1 } \Bigg[ \kP + \kQ \\
&~~~~- \frac{ ( \tA - \oA ) ( \kP - \kQ ) + 2 \sin^2 \phi_k ( \oA \kP + \tA \kQ ) }{ \sqrt{ ( \tA - \oA )^2 + 4 \oA \tA \sin^2 \phi_k } } \Bigg] \label{eq:p2}.
\end{split}
\end{equation}

Now we turn our attention to computing $P_k$. We can assume $d = 2$ since $P_k$ does not depend on the dimension. By using Clebsch-Gordan coefficients and the notations given by (\ref{eq:mu}), the projector in Eq. (\ref{eq:P0}) can be written as
\begin{equation}
\begin{split}
& \ktI \ot | \u \ra \la \u | \\
&= \sum_{ l = 0 }^{ N - 2 k } \sum_{ i = 0 }^{ N_1 } \sum_{ j = 0 }^{ M + N_2 }| \la \mu : \mu - l | \mu_1 : \mu_1 - i ; \mu_{ 02 } : \mu_{ 02 } - j \ra |^2  \\
&~~~\times | \mu_1 : \mu_1 - i \ra \la \mu_1 : \mu_1 - i | \ot | \mu_{ 02 } : \mu_{ 02 } - j \ra \la \mu_{ 02 } : \mu_{ 02 } - j | \label{eq:I}.
\end{split}
\end{equation}

In the following,
we fix the notation $| \ua \ra$ ( $| \da \ra$ ) to denote the vector in the space $\C^2$ whose weight is $\half$ ($- \half $). 
Without loss of generality, we can assume that $\rho_2$ is $| \ua \ra \la \ua |$.
Then, we can write
\begin{equation}
\uR = | \mu_{ 02 } : \mu_{ 02 } \ra \la \mu_{ 02 } : \mu_{ 02 } | 
=|\ua\ra\la \ua|^{\otimes M+N_2}
\label{eq:rho2}.
\end{equation}
Plugging equations (\ref{eq:I}) and (\ref{eq:rho2}) into Eq. (\ref{eq:P0}), we have
\begin{equation}
\begin{split}
&\kP = \sum_{ l = 0 }^{ N_1 - k } \la \mu_1 : \mu_1 - k - l | \1R | \mu_1 : \mu_1 - k - l \ra \\
&~~~\times | \la \mu : \mu - l | \mu_1 : \mu_1 - k - l ; \mu_{ 02 } : \mu_{ 02 } \ra |^2 \label{eq:P1half}.
\end{split}
\end{equation}
Moreover, converting the variable $l$ into $N_1 - k -l$, this can be written as
\begin{equation}
\begin{split}
&\kP= \sum_{ l = 0 }^{ N_1 - k }  \la \frac{ N_1 }{ 2 } : - \frac{ N_1 }{ 2 } + l | \1R | \frac{ N_1 }{ 2 } : - \frac{ N_1 }{ 2 } + l \ra \\
&\times | \la \frac{ N }{ 2 } - k : \frac{ N }{ 2 } - N_1 + l| \frac{ N_1 }{ 2 } : - \frac{ N_1 }{ 2 } + l ; \frac{ M + N_2 }{ 2 } : \frac{ M + N_2 }{ 2 } \ra |^2
\label{eq:P2}.
\end{split}
\end{equation}
 
We can calculate the Clebsch-Gordan coefficients \cite{Messiah} as 
\begin{equation}
\begin{split}
&| \la \frac{ N }{ 2 } - k : \frac{ N }{ 2 } - N_1 + l| \frac{ N_1 }{ 2 } : - \frac{ N_1 }{ 2 } + l ; \frac{ M + N_2 }{ 2 } : \frac{ M + N_2 }{ 2 } \ra |^2 \\
&~~=\frac{ ( N - 2 k + 1 ) ( N_1 - k ) ! ( N_1 - l) !  }{ ( N - k + 1 ) ! ( N_1 - k - l ) !  } \\
&~~~~~~~~~~~~~~~~~~\times \frac{( M + N_2 ) ! ( M + N_2 - k + l ) !}{( M + N_2 - k ) ! k ! l !} \label{CG}.
\end{split}
\end{equation}

Denoting $\rho_1=|\phi_1 \ra \la \phi_1|$,
we obtain $|\la \phi_1| \ua \ra|^2=q$.
Thus,
\begin{equation}
\begin{split}
& \la \frac{ N_1 }{ 2 } : - \frac{ N_1 }{ 2 } + l | \1R | \frac{ N_1 }{ 2 } : - \frac{ N_1 }{ 2 } + l \ra \\
& = \binom{ N_1 }{ l } | \la \phi_1 | \ua \ra |^{ 2 l } | \la \phi_1 | \da \ra |^{ 2 ( N_1 - l ) } \\
& = \binom{ N_1 }{ l } q^l ( 1 - q )^{ N_1 - l } \label{eq:rho1}.
\end{split}
\end{equation}

Therefore, we can write $P_k$ as
\begin{equation}
\begin{split}
&\kP \equiv \frac{ ( N - 2 k + 1 ) N_1 ! ( M + N_2 ) ! }{ ( N - k + 1 ) ! k ! } \\
&~~~\times \sum_{ l = 0 }^{ N_1 - k } \binom{ N_1 - k }{ l } \binom{ M + N_2 - k + l }{ l } q^l ( 1 - q )^{ N_1 - l } \label{eq:P3},
\end{split}
\end{equation}
where we have defined $0^0$ as $1$.

In the same way, one obtains
\begin{equation}
\begin{split}
&\kQ \equiv \frac{ ( N - 2 k + 1 ) ( M + N_1 ) ! N_2 ! }{ ( N - k + 1 ) ! k ! } \\
&~~~\times \sum_{ l = 0 }^{ N_2 - k } \binom{ M + N_1 - k + l }{ l } \binom{ N_2 - k }{ l }  q^l ( 1 - q )^{ N_2 - l } \label{eq:Q3}.
\end{split}
\end{equation}
\end{proof}

\section{Limit of the Minimum Averaged Error Probability}

When the both numbers of copies in System 1 and System 2 approach infinitely large, we have perfect knowledge to determine the states in the two systems.
In this limit, 
by using Eq. (\ref{eq:apeq}), 
the probability $\overline{ p }_{ M, N_1, N_1 } ( E_{ M, N_1, N_1 } )$ can be written as
\begin{equation}
\begin{split}
& \lim_{ N_1 \to \infty } \overline{ p }_{ M, N_1, N_1 } ( E_{ M, N_1, N_1 } ) \\
& ~~= \half \left[ 1 - 2 ( d - 1 ) \int_0^1 x \sqrt{ 1 - x^{ 2M } }( 1 - x^2 )^{ d - 2 } dx \right] \\
& ~~= \half \left[ 1 - ( d - 1 ) \int_0^1 \sqrt{ 1 - x^M }( 1 - x )^{ d - 2 } dx \right] \label{eq:limap},
\end{split}
\end{equation}
where we have defined $x = \frac{k}{n}$ and used the Euler-McLaurin summation formula.
The case of $d=2$ coincides with (18) in Santis et al \cite{Bagan},
and the case of $M=1$ coincides with (41) in A. Hayashi et al \cite{Hayashi}.

This result could be easily anticipated from the minimum error probability of the discrimination problem \cite{Helstorm}. 
Recall that the minimum error probability given $M$ identical copies is $\half \left[ 1 - \sqrt{ 1 - ( \Tr [ \rho_1 \rho_2 ] )^M } \right]$.
Assuming that $\rho_1$ and $\rho_2$ are distributed according to $\mu_{\Theta_d}$ independently, the average is given by
\begin{equation}
\begin{split}
& \int_{ \Theta_d } \int_{ \Theta_d } \half \left[ 1 - \sqrt{ 1 - ( \Tr [ \rho_1 \rho_2 ] )^M } \right] \mu_{ \Theta_d } ( d \rho_1 ) \mu_{ \Theta_d } ( d \rho_2 ) \\
& ~~= \half \left[ 1 - 2 ( d - 1 ) \int_0^{ \frac{ \pi }{ 2 } } \sqrt{ 1 - ( \cos \theta )^{ 2 M } } ( \sin \theta )^{ 2 d - 3 } \cos \theta d \theta  \right] \\
& ~~= \half \left[ 1 - ( d - 1 ) \int_0^1 \sqrt{ 1 - x^M }( 1 - x )^{ d - 2 } dx \right] \label{eq:avap}.
\end{split}
\end{equation}
Therefore, our optimal measurement can achieve the average performance of two state discrimination
under the limit $N_1=N_2 \to \infty$.

Next, we turn our attention to the complementary case, that is, the number of copies in System $0$ is infinitely large. 
By using Eq. (\ref{eq:ap4}), the minimum averaged error probability in this limit can be computed as
\begin{equation}
\lim_{ M \to \infty } \ap ( \E ) = \frac{ 1 }{ 2 \binom{ N_2 + d - 1 }{ d - 1 } } \label{eq:minfty}.
\end{equation}
Note that this result is independent of $N_1$. 

In this limit, we have perfect knowledge of the pure state $\rho$ in System $0$ and this problem is equal to distinguishing two states 
$\rho^{ \ot N_1 } \ot \hrho^{ \ot N_2 }$ and $\hrho^{ \ot N_1 } \ot \rho^{ \ot N_2 }$
in the composite system $12$.
This problem can be regarded as a generalization of state comparison \cite{comparison}.
As is shown in the following, the minimum error probability for these two states can be obtained with a POVM whose elements are 
$\{ E_1 = I_1 \ot ( I_2 - \rho^{ \ot N_2 } ),  E_2 = I_1 \ot \rho^{ \ot N_2 } \}$, 
where  $I_1$ ($I_2$) is the unit matrix on $( \C^d )^{ \ot N_1 }$ ($( \C^d )^{ \ot N_2 }$). 
Here, $E_1 ( E_2 )$ corresponds to the guess $\rho^{ \ot N_1 } \ot \hrho^{ \ot N_2 } (\hrho^{ \ot N_1 } \ot \rho^{ \ot N_2 } )$. 
We then can write the error probability as
\begin{equation}
\half \Tr [ \rho^{ \ot N_1 } \ot \hrho^{ \ot N_2 } E_2 ] + \half \Tr [ \hrho^{ \ot N_1 } \ot \rho^{ \ot N_2 } E_1 ]
= \half \left( \Tr [ \rho \hrho ] \right)^{ N_2 } \label{eq:comparison},
\end{equation}
Thus, the average is computed as follows:
\begin{equation}
\begin{split}
& \int_{ \Theta_d } \int_{ \Theta_d } \half \left( \Tr [ \rho \hrho ] \right)^{ N_2 } \mu_{ \Theta_d } ( d \rho ) \mu_{ \Theta_d } ( d \hrho ) \\
&~~ = ( d - 1 ) \int_0^{ \frac{ \pi }{ 2 } } ( \cos \theta )^{ 2 N_2 } ( \sin \theta )^{ 2 d - 3 } \cos \theta d \theta \\
&~~ = \frac{ d - 1 }{ 2 } \int_0^1 x^{ N_2 } ( 1 - x )^{ d - 2 } d x \\
&~~ = \frac{ 1 }{ 2 \binom{ N_2 + d - 1 }{ d - 1 } } \label{eq:comparison1}.
\end{split}
\end{equation}
Since this value is the same as the expression in Eq. (\ref{eq:minfty}),
the POVM $\{ E_1 = I_1 \ot ( I_2 - \rho^{ \ot N_2 } ),  E_2 = I_1 \ot \rho^{ \ot N_2 } \}$
realizes the optimal performance.

\section{Exponential Decreasing Rate of the Error Probability}

When every number of copies $N_1$, $N_2$ and $M$ is infinitely large, that is, the states in three systems are perfectly known, the error probability approaches zero unless $q = 1$. 
In order to treat the convergence speed,
we focus on the exponential decreasing rate of the error probability when the optimal POVM $E_{ M, N_1, N_1 }$ is applied.
For simplicity, we assume that the numbers $N_1$, $N_2$ of copies in Systems $1$, $2$
increase in proportion to the number $M$ of copies in System $0$.
When the proportional constant is given to be $\alpha>0$,
we have $N_1=N_2=\alpha M$.
Thus, using the real numbers
\begin{equation}
\begin{split}
&C_k \equiv \frac{ ( M + 2 \al M - 2 k + 1 ) ( \al M ) ! ( M + \al M ) ! }{ 2 ( M + 2 \al M - k + 1 ) ! k ! } \\
&~~\times \left[ 1 - \sqrt{ 1 - \left( \frac{ ( \al M ) ! ( M + \al M - k ) ! }{ ( M + \al M ) ! ( \al M -k ) ! } \right)^2 } \right] \label{eq:C},
\end{split}
\end{equation}
\begin{equation}
D_{ k, l } \equiv \binom{ \al M - k }{ l } \binom{ M + \al M - k + l }{ l } q^l ( 1 - q )^{ \al M - l } \label{eq:D},
\end{equation}
we can write from Eq. (\ref{eq:peq})
\begin{equation}
p_{ M, \al M, \al M } ( \rho_1, \rho_2, E_{ M, \al M, \al M } ) = \sum_{ k = 0 }^{ \al M }  \sum_{ l = 0 }^{ \al M - k } C_k D_{ k, l } \label{eq:pc}.
\end{equation}

The convergence speed is represented as
$\displaystyle \lim_{ M \to \infty } \frac{ - 1 }{ M } \log p_{ M, \al M, \al M } ( \rho_1, \rho_2, E_{ M, \al M, \al M } )$,
which can be deformed as
\begin{equation}
\begin{split}
& \lim_{ M \to \infty } \frac{ - 1 }{ M } \log p_{ M, \al M, \al M } ( \rho_1, \rho_2, E_{ M, \al M, \al M } ) \\
&~~ = \lim_{ M \to \infty } \frac{ - 1 }{ M } \log ( \sum_{ k = 0 }^{ \al M }  \sum_{ l = 0 }^{ \al M - k } C_k D_{ k, l } ) \\
&~~ = \lim_{ M \to \infty } \frac{ - 1 }{ M } \log ( \max_{ 0 \le k \le \al M } \max_{ 0 \le l \le \al M - k } C_k D_{ k, l } ) \label{eq:pc1}.
\end{split}
\end{equation}
Moreover, using the approximation formula $\sqrt{ 1 - x } \thickapprox 1 - \half x $( when $x \ll 1 $)and the Stirling approximation $n ! \thickapprox n^n e^{ - n } \sqrt{ 2 \pi n }$, one obtains
\begin{equation}
\begin{split}
&\lim_{ M \to \infty } \frac{ - 1 }{ M } \log p_{ M, \al M, \al M } ( \rho_1, \rho_2, E_{ M, \al M, \al M } ) \\
&~~= \min_{ 0 \le \be \le \al } \min_{ 0 \le \ga \le \al - \be } h ( \be, \ga ) \label{eq:pc2},
\end{split}
\end{equation}
where we have defined as $\be \equiv \frac{ k }{ M }, \ga \equiv \frac{ l }{ M }$ and
\begin{equation}
\begin{split}
&h ( \be, \ga ) \equiv ( \al - \be - \ga ) \log ( \al - \be - \ga ) \\
&~- ( 1 + \al - \be + \ga ) \log ( 1 + \al - \be + \ga )+ ( \al - \be ) \log ( \al - \be ) \\ 
&~- ( 1 + \al - \be ) \log ( 1 + \al - \be )- 3 \al \log \al \\
&~+ ( 1 + \al ) \log ( 1 + \al ) + \be \log \be \\
&~+ ( 1 + 2 \al - \be ) \log ( 1 + 2 \al - \be ) \\
&~+ 2 \ga \log \ga - \ga \log q - ( \al - \ga ) \log ( 1 - q ) \label{eq:h}.
\end{split}
\end{equation}

There is the unique root of $\frac{\partial h}{\partial \be} = \frac{\partial h}{\partial \ga} = 0$ in the range $0 < \be < \alpha$, $0 < \ga < \alpha - \be$ and we use $( \be_1, \ga_1 )$ to denote it. 
These can be calculated as
\begin{align}
&\ga_1 = \frac{ q ( \al - 1 ) + \sqrt{ q^2 ( \al - 1 )^2 + 4 q \al } }{ 2 }  \label{eq:ga}, \\
&\be_1 = \frac{ ( 2 \al + 1 ) - \sqrt{ ( 2 \al + 1 )^2 - 4 ( \al^2 - \al \ga_1 ) } }{ 2 } \label{eq:be}.
\end{align}
One can agree that $h ( \be_1, \ga_1 )$ is the minimum of the function $h$ in the range $0 < \be < \alpha$, $0 < \ga < \alpha - \be$
due to the following equations:
\begin{equation}
\begin{split}
&~~~~~~~~~\frac{\partial^2 h}{\partial \be^2} ( \be, \ga ) > 0, \\
&~~~~~~~~~\frac{\partial^2 h}{\partial \ga^2} ( \be, \ga ) > 0, \\
&~~\displaystyle \lim_{ \be \to 0 } \frac{\partial h}{\partial \be}( \be, \ga ) = \lim_{ \ga \to 0 } \frac{\partial h}{\partial \ga} ( \be, \ga ) = - \infty , \\
&\displaystyle \lim_{ \be \to \al - \ga } \frac{\partial h}{\partial \be}( \be, \ga ) = \lim_{ \ga \to \al - \be } \frac{\partial h}{\partial \ga} ( \be, \ga ) = + \infty , \label{eq:conditions}
\end{split}
\end{equation}

\begin{figure}[t]
\begin{center}
\scalebox{1.0}{\includegraphics[scale=0.8]{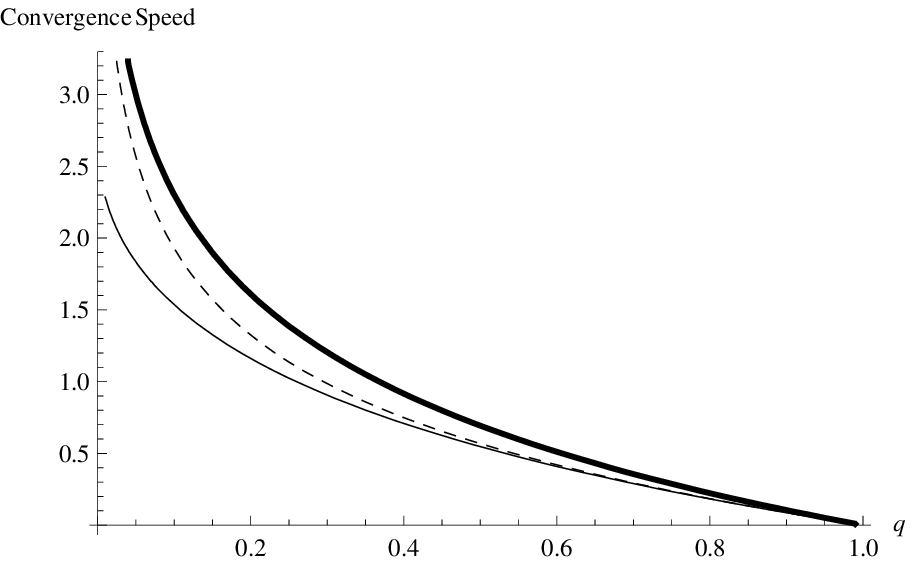}}
\end{center}

\caption{(normal line) $h ( \be_1, \ga_1 )$, (dashed line) $- \log q - \frac{ 1 - q }{ q ( \al - 1) + 2 }$, (thick line) $- \log q$ for $\alpha$ = 5}\label{f1}
\end{figure}

\begin{figure}[t]
\begin{center}
\scalebox{1.0}{\includegraphics[scale=0.8]{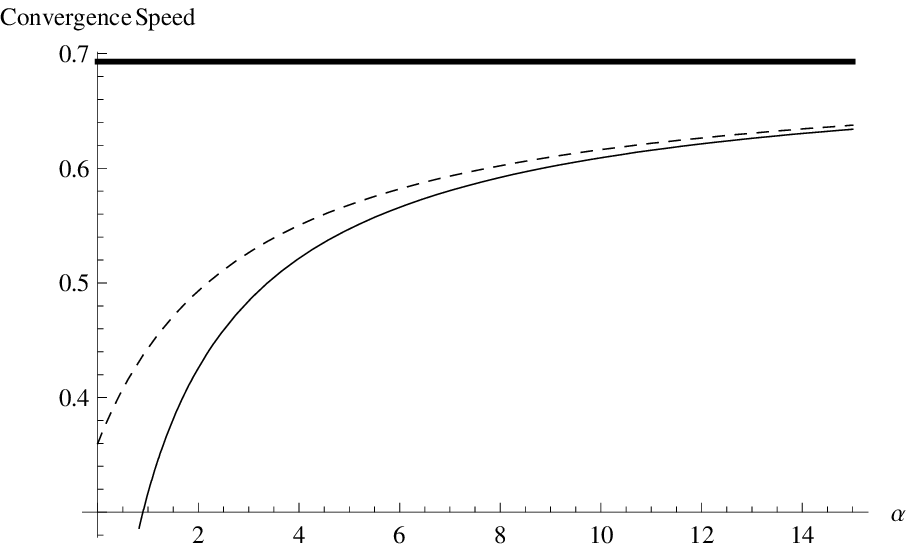}}
\end{center}
\caption{(normal line) $h ( \be_1, \ga_1 )$, (dashed line) $- \log q - \frac{ 1 - q }{ q ( \al - 1) + 2 }$, (thick line) $- \log q$ for $q = 0.5$}\label{f2}
\end{figure}

When $\alpha$ is sufficient large, we can write
\begin{equation}
h ( \be_1, \ga_1 ) = - \log q - \frac{ 1 - q }{ q ( \al - 1) + 2 } + O ( \frac{ 1 }{ \al^2 } ) \label{eq:h1}.
\end{equation}
In fact, 
as is numerically demonstrated in Figs. \ref{f1} and \ref{f2},
$- \log q - \frac{ 1 - q }{ q ( \al - 1) + 2 }$ well approximates 
$h ( \be_1, \ga_1 )$ when $\alpha$ is large.
Therefore, we obtain the convergence speed of the error probability,
\begin{equation}
\begin{split}
&\lim_{ M \to \infty } \frac{ - 1 }{ M } \log p_{ M, \al M, \al M } ( \rho_1, \rho_2, E_{ M, \al M, \al M } ) \\ 
&~= - \log \Tr [ \rho_1 \rho_2 ] - \frac{ 1 - \Tr [ \rho_1, \rho_2 ]}{ \Tr [ \rho_1 \rho_2 ] ( \al - 1 ) + 2 } + O ( \frac{ 1 }{ \al^2 } ) \label{eq:pc3}.
\end{split}
\end{equation}

In the discrimination problem of two pure states \cite{Helstorm}, 
when the number of copies of the state to be identified is infinitely large, 
the convergence speed is given by
\begin{equation}
\lim_{ M \to \infty } \frac{ - 1 }{ M } \log \half \left[ 1 - \sqrt{ 1 - ( \Tr [ \rho_1 \rho_2 ] )^M } \right]
= - \log \Tr [ \rho_1 \rho_2 ] \label{eq:pcu}.
\end{equation}
This is called the quantum Chernoff bound \cite{Chernoff} and equal to the limit of Eq. (\ref{eq:pc3}) as $\alpha \rightarrow \infty$.
This fact means that
the performance of our optimal POVM is close to 
that of the optimal POVM in the sense of quantum state discrimination.

\section{conclusions}

We have studied the changepoint problem in the quantum setting, where our task is to choose the correct changepoint between two candidates. This problem is equal to discriminating two unknown general states when multiple copies of the state are provided. We have obtained the minimum averaged error probability, Eq. (\ref{eq:ap4}). 
Our result of special cases coincides with
the results by \cite{Bagan} and \cite{Hayashi}. 
However, when arbitrary numbers of copies of general pure states are given, we have calculated it for the first time. 
Moreover, we have first calculated the non-averaged error probability Eq.(\ref{eq:p1}). This depends on the inner product. As could be anticipated, when the numbers of copies of candidates are infinitely large, we recover the average of the usual discrimination problem. We have also paid attention to the exponential decreasing rate and shown the convergence rate of the non-averaged error probability approaches the quantum Chernoff bound.


\section*{Acknowledgment}
The authors thank Professor Leong-Chuan Kwek for pointing out the
importance of the changepoint problem.
They are grateful for Dr. Gen Kimura and Mr. Wataru Kumagai
to helpful discussion concerning qunautm information.
The authors are 
partially supported by a Grant-in-Aid for Scientific Research in the
Priority Area ``Deepening and Expansion of Statistical Mechanical
Informatics (DEX-SMI),'' No. 18079014 and a MEXT Grant-in-Aid for
Young Scientists (A) No. 20686026.
The Centre for Quantum Technologies is funded by the
Singapore Ministry of Education and the National Research Foundation
as part of the Research Centres of Excellence programme.

\appendix
\section{Calculation of Wigner's 6j-function}

Let us consider the calculation of the Wigners 6j-function $\begin{Bmatrix} a & b & c \\ d & e & f \end{Bmatrix}$. Let us define some notations as
\begin{equation}
\begin{split}
& \al_1 \equiv a + b + d + e, \al_2 \equiv a + c + d + f, \al_3 \equiv b + c + e + f, \\
& \be_1 \equiv a + b + c, \be_2 \equiv a + e + f, \\
& \be_3 \equiv b + d + f, \be_4 \equiv c + d + e, \label{eq:6j1}
\end{split}
\end{equation}
and let us define $A_1, A_2$ and $A_3$ to be the smallest, middle, and largest values of $\alpha_1, \alpha_2$ and $\alpha_3$ and $B_1, B_2, B_3$ and $B_4$ to be the smallest, second smallest, second largest, and largest values of $\be_1, \be_2, \be_3$ and $\be_4$. When $B_4 = A_1$, from the formula in \cite{6j} we can calculate as
\begin{equation}
\begin{split}
&\begin{Bmatrix} a & b & c \\ d & e & f \end{Bmatrix}
=( - 1 )^{ B_4 } \left[ \prod_{ i = 1 }^{ B_4 - B_3 } \frac{ ( B_3 + 1 + i ) ( A_3 - A_1 + i ) }{ ( A_3 - B_1 + i ) ( A_2 - B_2 + i ) } \right]^{ \half } \\
&~~\times \Bigg[ \frac{ 1 }{ ( B_1 + 1 ) ( B_2 + 1 ) } \\
&~~\times \prod_{ i = 1 }^ { A_2 - A_1 } \frac{ ( A_1 - B_1 + i ) ( B_4 - B_2 + i ) ( B_4 - B_3 + i ) }
{ ( A_1 + B_1 - A_2 + i ) ( A_1 + B_2 - A_2 + i ) i } \Bigg]^{ \half } \label{eq:atheorem}.
\end{split}
\end{equation}

We now compute the Wignerfs 6-j funcion $\begin{Bmatrix} \mu_1 & \mu_0 & \mu_{ 01 } \\ \mu_2 & \mu  & \mu_{ 02 } \end{Bmatrix}$ in Eq.(\ref{eq:ip}).

\subsection*{In case $0 \le k \le M$}

In this case, we can write as
\begin{equation}
\begin{split}
& A_1 = N - k, A_2 = N, A_3 = M + N - k \\
& B_1 = M + N_1 , B_2 = M + N_2, B_3 = N - k, B_4 = N - k  \label{eq:6j3}.
\end{split} 
\end{equation}
Plugging these into Eq. (A2), one obtains
\begin{equation}
\begin{split}
&\begin{Bmatrix} \mu_1 & \mu_0 & \mu_{ 01 } \\ \mu_2 & \mu  & \mu_{ 02 } \end{Bmatrix} = ( - 1 )^{ N - k } \Bigg[ \frac{ 1 }{ ( M + N_1 + 1 ) ( M + N_2 + 1 ) } \\
&~~\times \prod_{ i = 1 }^k \frac{ ( N_1 - k + i ) ( N_2 - k + i ) }{ ( M + N_1 - k + i ) ( M + N_2 - k + i ) } \Bigg]^{ \half } \label{eq:6f3half}.
\end{split}
\end{equation}
This is deformed as
\begin{equation}
\begin{split}
&\begin{Bmatrix} \mu_1 & \mu_0 & \mu_{ 01 } \\ \mu_2 & \mu  & \mu_{ 02 } \end{Bmatrix} \\
&= \frac{ ( -1 )^{ \mu_0 + \mu_1 + \mu_2 + \mu } }{ \sqrt{ ( 2 \mu_{ 01 } + 1 ) ( 2 \mu_{ 02 } + 1 ) } }
\sqrt{ \frac{ \binom{ N_1 }{ k } \binom{ N_2 }{ k } }{ \binom{ M + N_1 }{ k } \binom{ M + N_2 }{ k } } } \label{eq:6j4}.
\end{split}
\end{equation}

\subsection*{In case $M + 1 \le k \le N_1$}

In this case, we can write as
\begin{equation}
\begin{split}
& A_1 = N - k, A_2 = M + N - k, A_3 = N \\
& B_1 = M + N_1 , B_2 = M + N_2, B_3 = N - k, B_4 = N - k \label{eq:6j5}.
\end{split} 
\end{equation}
Plugging these into Eq. (A2), one obtains
\begin{equation}
\begin{split}
\begin{Bmatrix} \mu_1 & \mu_0 & \mu_{ 01 } \\ \mu_2 & \mu  & \mu_{ 02 } \end{Bmatrix}
&~= ( - 1 )^{ N - k } \Bigg[ \frac{ 1 }{ ( M + N_1 + 1 ) ( M + N_2 + 1 ) } \\
&~~\times \prod_{ i = 1 }^M \frac{ ( N_1 - k + i ) ( N_2 - k + i ) }{ ( N_1 + i ) ( N_2 + i ) } \Bigg]^{ \half } \label{eq:6j6}.
\end{split}
\end{equation}
Since $M + 1 \le k$, one has
\begin{equation}
\frac{ \binom{ N_1 }{ k } \binom{ N_2 }{ k } }{ \binom{ M + N_1 }{ k } \binom{ M + N_2 }{ k } }
 = \prod_{ i = 1 }^M \frac{ ( N_1 - k + i ) ( N_2 - k + i ) }{ ( N_1 + i ) ( N_2 + i ) } \label{eq:6j7}.
\end{equation}

Thus, 
\begin{equation}
\begin{split}
&\begin{Bmatrix} \mu_1 & \mu_0 & \mu_{ 01 } \\ \mu_2 & \mu  & \mu_{ 02 } \end{Bmatrix} \\
&= \frac{ ( -1 )^{ \mu_0 + \mu_1 + \mu_2 + \mu } }{ \sqrt{ ( 2 \mu_{ 01 } + 1 ) ( 2 \mu_{ 02 } + 1 ) } }
\sqrt{ \frac{ \binom{ N_1 }{ k } \binom{ N_2 }{ k } }{ \binom{ M + N_1 }{ k } \binom{ M + N_2 }{ k } } } \label{eq:6j8}.
\end{split}
\end{equation}
also holds.
\vspace{120pt}

\end{document}